\documentclass[aps,prb,superscriptaddress,preprintnumbers,
showpacs,legalpaper,twoside,twocolumn,amsmath,amssymb]{revtex4}
\usepackage{graphicx}     
\begin{document}
\title{Study of Short-distance Spin and Charge Correlations and Local Density-of-States \\
in the CMR regime of the One-Orbital Model for Manganites}
\author{Rong Yu}
\affiliation{Department of Physics and Astronomy, University of
Tennessee, Knoxville, TN 37996, USA} \affiliation{Materials Science
and Technology Division, Oak Ridge National Laboratory, Oak Ridge,
TN 32831, USA}
\author{Shuai Dong}
\affiliation{Department of
Physics and Astronomy, University of Tennessee, Knoxville, TN 37996, USA}
\affiliation{Materials Science and Technology Division, Oak
Ridge National Laboratory, Oak Ridge, TN 32831, USA}
\affiliation{Nanjing National Laboratory of Microstructures, Nanjing
University, Nanjing 210093, China}
\author{Cengiz \c{S}en}
\affiliation{Department of Physics, University of Cincinnati,
Cincinnati, OH 45221, USA}
\author{Gonzalo Alvarez}
\affiliation{Computer Science and Mathematics Division and Center
for Nanophase Materials Science, Oak Ridge National Laboratory, Oak
Ridge, TN 37831, USA}
\author{Elbio Dagotto}
\affiliation{Department of Physics and Astronomy, University of
Tennessee, Knoxville, TN 37996, USA} \affiliation{Materials Science
and Technology Division, Oak Ridge National Laboratory, Oak Ridge,
TN 32831, USA}


\begin{abstract}
The metal-insulator transition, and the
associated magnetic transition, in the colossal magnetoresistance (CMR) regime of
the one-orbital model for
manganites is here studied using Monte Carlo (MC) techniques. Both cooperative oxygen lattice distortions
and a finite superexchange coupling among the $t_{\rm 2g}$ spins are included in our investigations.
Charge and spin
correlations are studied. In the CMR regime, a strong
competition between the ferromagnetic metallic and
antiferromagnetic charge-ordered insulating states is observed.
This competition is shown to be important to understand the resistivity
peak that appears near the critical temperature. Moreover, it is argued that the system is dynamically
inhomogeneous, with short-range charge and spin
correlations that slowly evolve with MC time, producing the glassy characteristics of the CMR
state.
The local density-of-states (LDOS) is also investigated, and
a pseudogap (PG) is found to exist in the CMR temperature range. The
width of the PG in the LDOS is calculated
and directly compared with recent
scanning-tunneling-spectroscopy (STS) experimental results. The agreement
between our calculation and the experiment suggests that the
depletion of the conductance at low bias observed experimentally is a reflection
on the existence of a PG in the LDOS spectra, as opposed to a hard gap.
The apparent homogeneity observed via STS techniques could be caused by the
slow time characteristics of this probe. Faster experimental methods should unveil a
rather inhomogeneous state in the CMR regime, as already observed in neutron scattering experiments.
\end{abstract}
\maketitle

\section{Introduction}
The colossal magnetoresistance effect in manganites continues
attracting considerable interest in both condensed matter
physics and material
science.\cite{RaoBook98,TokuraBook00,DagottoReview01,DagottoScience05,SalamonJaime01}
Clarifying the origin of the CMR effect is expected to
fundamentally contribute to our
understanding of the intrinsic complexity observed in a variety of
transition metal oxides. In addition, these
compounds are receiving much attention
for their potential application in new generations of spintronics devices.\cite{oxide-electronics}
Manganites present a very rich phase diagram with a variety of
competing states, giving rise to the exotic behavior found in
these compounds. To explain the CMR effect, originally a double exchange (DE)
model was proposed.\cite{Zener51} However, it was shown later
that the DE mechanism itself could not explain the large changes in the resistivity
with increasing magnetic fields that were observed
experimentally.\cite{Millisetal95, Calderonetal99} Thus, other
couplings, such as the electron-phonon interactions, must be taken into
account.\cite{Millisetal96} Subsequent investigations have shown that
the existence of insulating states competing with the DE
induced ferromagnetic metal, and the concomitant emergence of nanometer-scale
phase separation, could be the key ingredients to understand the CMR
effect.\cite{Yunokietal98,Moreoetal99,DagottoBook}

\begin{figure}[h]
\begin{center}
\includegraphics[
bbllx=0pt,bblly=30pt,bburx=419pt,bbury=534pt,%
     width=75mm,angle=0]{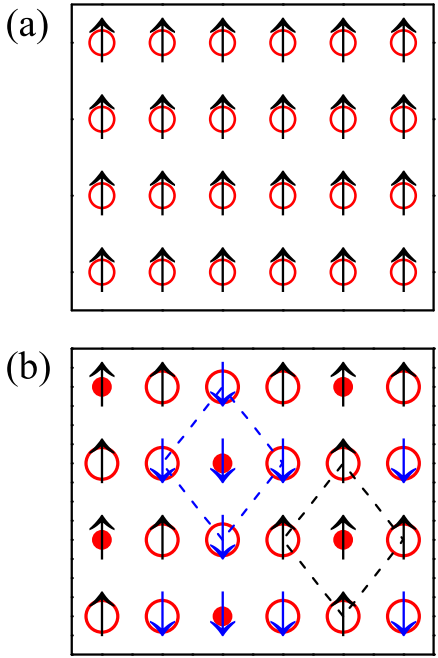}
\caption{(color online) The two competing ground states in the
one-orbital model at electronic density $n$=0.75, obtained by varying the
superexchange coupling $J_{\rm AF}$, as explained in Ref.~\onlinecite{Sen07} and in
the text. The FM metallic
state is in (a) and the AF/CO insulating state in (b). The arrows denote
the classical $t_{\rm 2g}$ spins and the open circles are proportional to the
electronic density. The full circles in (b) simply denote the location of the holes.
The dashed lines in (b) define ``spin blocks'', that are discussed later
in the text.} \label{F.GSsnap}
\end{center}
\end{figure}

After the phase separation idea was proposed, a variety of theoretical studies have
been reported, including calculations using simplified spin
models and random resistor
networks.\cite{Burgyetal01,Mayretal02,Burgyetal04}
More realistic one-orbital and two-orbital models, including
electron-phonon coupling and quenched
disorder,\cite{Vergesetal02,KumarMajumdar05,KumarMajumdar06,salafranca,Senetal06} have
also been investigated.
In several of these calculations a CMR behavior of the resistivity vs. temperature
has been observed. The
ferromagnetic (FM) metallic phase is commonly accepted as one
of the competing states necessary for CMR. The other competitor
should be an antiferromagnetic (AF) charge-ordered (CO)
insulating state. But for technical reasons related with the complexity of the numerical simulations,
the CMR arising from the competition between
the FM metallic and the AF/CO insulating states has not been
discussed in most of the efforts mentioned above. However, very recently \c{S}en {\it et
al.}~\cite{Sen07} carefully analyzed the influence of the superexchange coupling $J_{\rm AF}$
between the localized $t_{\rm 2g}$ electrons on the physical properties of both the one- and
two-orbital models including cooperative oxygen lattice distortions. Two competing ground
states were discovered at low temperatures: a FM metallic phase and an
AF/CO insulating phase (for the charge and spin arrangement in these states see
Fig.~\ref{F.GSsnap}).
These two phases are separated by first-order transitions at low temperatures.
A thermal transition from either phase to a paramagnetic (PM)
insulator at high temperature was also observed.
Close to the bi-critical point in the phase diagram,
the CMR effect in the resistivity was found.\cite{bicritical} This is a location where
enhanced phase competition is observed. The CMR effect has been
shown to be deeply connected with the existence of
short-distance charge correlations above the
critical temperature of the thermal transition.\cite{Sen07}

In this paper,  the recent effort by \c{S}en {\it et al.}\cite{Sen07} is further expanded, focusing on the
analysis of the phase competition in the thermal metal-insulator (MI) transition of the one-orbital model. We
have observed that the development of  short-range spin-spin correlations, with characteristics similar to those in
the AF/CO competing state, is important to understand the CMR effect. This result supplements
the previous investigations that focused on the charge sector exclusively.\cite{Sen07}
The systems without quenched
disorder that we study in this work are shown to be statistically homogeneous when averaged over very long MC times.
However, charge localization together with robust short-distance correlations of charges and spins
are observed at short MC times. This leads us to argue that a {\it dynamical} inhomogeneity occurs in
manganites in cases where the strength of the quenched disorder is weak (namely for materials close to
the clean limit).

A second purpose of our present effort is to use numerical
techniques to calculate the LDOS. Since the one-orbital model has
been shown to successfully describe the behavior of the resistivity
in the CMR regime,\cite{Sen07} this approach should be able to
explain other properties in the same temperature range. The LDOS is
important to contrast theory with STS experiments, which are useful
to probe local electronic structures in a variety of complex
materials, possibly providing information about phase separation in
manganites.\cite{Fathetal99,Renneretal02} In fact, the derivative of
the experimentally obtained current-voltage curve is directly
related to the LDOS. Our calculations are even more relevant in view
of recently reported puzzling STS results. In
Refs.~\onlinecite{Seiroetal07, Singhetal07}, a depletion of the
conductance at low bias was observed in the {\it low} temperature
metallic phase of both $\rm La_{0.77}Ca_{0.23}MnO_3$ (LCMO) and $\rm
La_{0.350}Pr_{0.275}Ca_{0.375}MnO_3$ (LPCMO) films, implying the
existence of a gapped quasi-particle excitation spectra in these
materials. No coexistence of insulating and metallic regions was
observed in the experiment, in apparent disagreement with existing
theories. These results are very surprising since a gapped spectrum
cannot lead to a metallic state. Singh {\it et
al.}\cite{Singhetal07} proposed that the surface of LPCMO is
different from the bulk: at the surface the AF/CO state is stable,
while in the bulk a coexistence of metallic and insulating regions
dominates, as shown in  other experiments. We do not disagree with
this proposal, and for LPCMO it may well be the solution of the
puzzle. For LCMO, the same hypothesis may work, but here we propose
another alternative. In our calculations reported below, a PG in the
LDOS (as opposed to a gap) is found in the entire temperature range
where the CMR occurs in the one-orbital model. This PG feature is
known to be a signature of doped manganites and results from
mixed-phase tendencies.\cite{Moreoetal99b,PGexperiments} We have
investigated the width of the PG and compare it with the
experimentally determined gap width. We have observed a very good
agreement between our calculation and the experimental results. This
provides an alternative way to understand the STS data for LCMO: the
experimentally observed gap could arise from a PG in the LDOS, not
from a gap. Moreover, within this picture, the apparent spatial
homogeneity observed in the STS experiment can be explained as a
time-averaged property. The system is actually dynamically
inhomogeneous, which could be experimentally detected if the time
scale of the probing technique is comparable to the characteristic
time of the dynamics of the system. Fast measurements (using, e.g.,
neutron scattering) have already shown evidence of ``correlated
polarons'' in LCMO\cite{correlated-polarons} in contradiction with
LCMO STS experiments. Our calculations show that this is to be
understood based on the dynamical nature of the CMR state: a
homogeneous gap detected in STS is not in contradiction with other
faster measurements, to the extent that it is understood that the
gap is  actually a PG in the LDOS.

This paper is organized as follows: in Sec. II, we briefly introduce
the one-orbital model and the details of the MC technique
that we used. In Sec. III, we focus
on investigating the competition between the FM metallic and the
AF/CO insulating states. Spin correlations are shown to play an
important role in the system, particularly in the CMR regime.
Moreover, our investigations in this section show that
the phase competition in the CMR regime is
dynamic, at least within the time evolution generated by the MC procedure, which
is based on local interactions. In Sec. IV, a detailed study of the PG in the LDOS of the
one-orbital model is provided. We compare our results with those in
the recent STS experiments, and show that their observations can be
explained within our model. Conclusions are in Sec. V.


\section{Theoretical model}
In this study, the properties of the one-orbital model for manganites
are analyzed using numerical techniques. Although a two-orbital model
with Jahn-Teller couplings is more realistic, it is also more difficult to
study computationally. In fact, using the exact diagonalization procedure
for the fermionic sector, a simulation using two orbitals costs $2^4$=16
times more CPU time than using only one orbital. Since it has been extensively
shown that the one-orbital model captures the physics of phase competition
in manganites,\cite{DagottoBook} being restricted to one orbital is not a drastic approximation.
Moreover, the current effort builds upon the previous
results obtained in Ref.~\onlinecite{Sen07}, where the one-orbital model
was also considered. Then, for a proper comparison, it is imperative to use the
same model. The Hamiltonian reads
\begin{eqnarray}\label{E.Ham1b}
H &=&-t\sum_{\langle ij\rangle,\alpha} (d^\dagger_{i,\alpha}
d_{j,\alpha} +h.c.) -J_{\rm H} \sum_{i,\alpha,\beta}
d^{\dagger}_{i,\alpha} \mathbf{\sigma}_{\alpha,\beta} d_{i,\beta}
\cdot \mathbf{S}_i \nonumber \\
& & + J_{\rm AF}\sum_{\langle
ij\rangle}\mathbf{S}_i\cdot\mathbf{S}_j -\lambda
t\sum_{i,\delta,\alpha} (u_{i,-\delta} - u_{i,\delta})
d^\dagger_{i,\alpha} d_{i,\alpha} \\
& &+ t\sum_{i,\delta} u^2_{i,\delta} +\sum_{i,\alpha}(\Delta_i -
\mu)n_{i,\alpha}, \nonumber
\end{eqnarray}
where $d_{i,\alpha}$ and $d^\dagger_{i,\alpha}$ are the fermionic
annihilation and creation operators of $e_{\rm g}$ electrons, acting on site $i$ and with $z$-axis spin
projection $\alpha$. $\{ \mathbf{\sigma} \}$ denote the Pauli matrices. $\mathbf{S}_i$
represents the $t_{\rm 2g}$ spin degrees-of-freedom localized
at site $i$. This spin is assumed to be classical, a widely
used approximation also employed in Ref.~\onlinecite{Sen07}.
The first two terms in Eq.~\ref{E.Ham1b} correspond to the ordinary
DE model. The third term introduces AF
interactions between the localized $t_{\rm 2g}$ spins. Although the coupling
regulating the strength of this term is considered
weak in magnitude,\cite{DagottoBook} previous work has shown that its presence is crucial to
stabilize the AF/CO states that compete with the FM ground state
resulting from the double-exchange mechanism.\cite{JAFcrucial} The next term in the
Hamiltonian contains the interaction between the mobile electrons
and the cooperative phonons ($\delta$ runs over the spatial directions).
As in Ref.~\onlinecite{Sen07} and several other investigations,
the oxygen lattice displacements $u_{i,\delta}$ with respect to the equilibrium position
are considered to be classical. These displacements are located at the links
of a two-dimensional square lattice. The oxygen lattice distortions
induced by the electron-phonon interaction are further balanced by
an elastic energy proportional to $u^2_{i,\delta}$. The last
term in the Hamiltonian Eq.~\ref{E.Ham1b} contains the quenched on-site
disorder that simulates the effect of chemical doping in real manganites.
As in Ref.~\onlinecite{Sen07} and other publications, this disorder should enter via a
randomly chosen set of values for the on-site energy $\Delta_i$  at every site.
However, although it has been shown that the quenched disorder can enhance
substantially the magnitude of the CMR effect,\cite{quenched}
it has been observed~\cite{Sen07}
that its presence would not change
the physical picture qualitatively. If quenched disorder is ignored, the only price to pay
is the need to tune couplings to be very close to the first-order FM-AF/CO transition
in order to observe CMR physics.\cite{Sen07} Then, for numerical simplicity, here we consider only
the clean limit CMR, i.e., $\Delta_i=0$ and focus on the immediate vicinity of the FM-AF/CO
transition in parameter space. As a consequence, our conclusions below regarding the dynamical character of the
inhomogeneities will then be of much more relevance to ``clean'' manganites than ``dirty'' ones.
Finally, $\mu$ is the chemical potential
and $n_{i,\alpha}$ is the number operator.

In the present calculation, and also for
simplicity, the limit $J_{\rm H}\to \infty$ is considered. This is another of the several
well-accepted
approximations widely used in previous studies of
theoretical models for manganites. This limit preserves the essential
physics of the CMR effect. According to this approximation, the spin of the $e_{\rm g}$
electrons is always parallel to the direction of the spin of the $t_{\rm 2g}$ electrons.
Hence, the Hamiltonian is reduced to
\begin{eqnarray}\label{E.HamJHinf}
H &=& -t\sum_{\langle ij\rangle}(\Omega_{ij}c^\dagger_i c_j+h.c.) +
J_{AF}\sum_{\langle ij\rangle}\mathbf{S}_i\cdot\mathbf{S}_j \nonumber \\
& &-\lambda t\sum_{i,\delta} (u_{i,-\delta} - u_{i,\delta})
c^\dagger_{i} c_{i} \\
& & + t\sum_{i,\delta} u^2_{i,\delta} -\sum_{i,\alpha} \mu
c^\dagger_{i} c_{i}, \nonumber
\end{eqnarray}
where $c^\dagger_i$ creates an $e_{\rm g}$ electron with spin parallel to
the localized $t_{\rm 2g}$ spin at site $i$. The classical localized spin can
be parametrized as $\mathbf{S}_i=(\sin \theta_i \cos \phi_i, \sin
\theta_i \sin \phi_i, \cos \theta_i)$ in spherical coordinates. The
influence of the infinite Hund coupling is reflected in the electron
hopping term that now carries a Berry phase:\cite{DagottoReview01}
\begin{equation}\label{E.Omega_ij}
\Omega_{ij} = \cos \frac{\theta_i}{2} \cos \frac{\theta_j}{2} + \sin
\frac{\theta_i}{2} \sin \frac{\theta_j}{2} e^{i(\phi_i-\phi_j)}.
\end{equation}
The Hamiltonian Eq.~\ref{E.HamJHinf} is solved here via a standard
combination of exact diagonalization of the fermionic sector and a
MC evolution of the classical spin and oxygen lattice distortion
degrees-of-freedom. At each MC step, the quadratic fermionic portion in the
Hamiltonian is diagonalized numerically, using library subroutines,
for a given configuration of classical
spins and phonons. A new set of configurations for the
next MC step is then generated following the Metropolis algorithm.
This method has been successfully used in previous studies and
details can be found in Refs.~\onlinecite{Yunokietal98,DagottoReview01} and
\onlinecite{DagottoBook}. Physical quantities can be easily
calculated as MC averages. For instance, at each MC step, the LDOS is evaluated
using the eigenvalues and eigenvectors that arise from the
diagonalization.\cite{DagottoBook} The resistivity $\rho$ is
obtained from the inverse of the average conductivity $\sigma$. In
two dimensions (2D), where our effort focuses on, it can be shown that
$\sigma = G$, where $G$ is the conductance. This conductance is
calculated using the Kubo
formula.\cite{Verges99} In this effort, 2D systems with
charge density $n$=$0.75$ are studied for a proper comparison with the
results in Ref.~\onlinecite{Sen07} that were obtained at the same density and dimensionality. The typical
clusters analyzed here are of size $8\times 8$ and $12\times 12$. If not
specified, simulations start with a random configuration of spins and lattice displacements.
To overcome the metastabilities caused by the
dynamical inhomogeneity that appears in this study (discussed in detail in the following
sections), very long MC runs were needed. In practice, up to $10^5$ MC steps for
thermalization were used, followed by  $5\times10^4$ steps for measurements. This
was needed on the $8\times 8$ lattice to obtain
statistically homogeneous results. For the same purpose, $10^4$ steps
for thermalization and another $10^4$ steps for measurements
were used for the $12\times12$ lattice.


\section{Competition between Ferromagnetic and Charge-Ordered
Antiferromagnetic states}

It has been shown~\cite{Sen07} that the model Eq.~\ref{E.HamJHinf} has
two competing states at
low temperatures, at the $n$=0.75 density
considered here.
The ground state is a FM metal if $J_{\rm AF}$ is zero or very small.
However, upon increasing $J_{\rm AF}$ an AF/CO insulator is reached via a
first-order phase transition. In this insulating phase, both charges and spins are arranged
in a non-homogeneous, but regular, pattern.
This is significantly different from the FM state,
where both charge and spin are uniformly distributed, namely they are the same at every site.
At high temperatures, the system loses long-range order and it is in a
paramagnetic (PM) phase.
The most striking feature found in this model~\cite{Sen07} is that
close to the phase boundary between the FM and AF/CO states,
the cooling down of the system from the high-temperature PM phase causes
drastic changes in
the resistivity $\rho$. First, $\rho$ increases with decreasing
temperature and develops a prominent peak. It then drops rapidly (about 2-3
orders of magnitude) when the system enters into the FM state. The
resistivity peak can be largely suppressed by small magnetic fields,
showing the so-called CMR effect.\cite{Sen07} While it is clear that
the AF/CO state of the one-orbital model is not exactly as found in experiments,
which are dominated by CE states that also present orbital order, the
promising results for $\rho$ vs. temperature show
that the key qualitative features of the CMR effect do appear
in the one-orbital model.

In this section, we will focus on the MI transition that causes the
CMR, and show how this CMR effect is connected to the
competition FM vs. AF/CO.


\subsection{Metal-Insulator Transition and the CMR Effect}
\begin{figure}[h]
\begin{center}
\includegraphics[
bbllx=120pt,bblly=10pt,bburx=750pt,bbury=600pt,%
     width=70mm,angle=0]{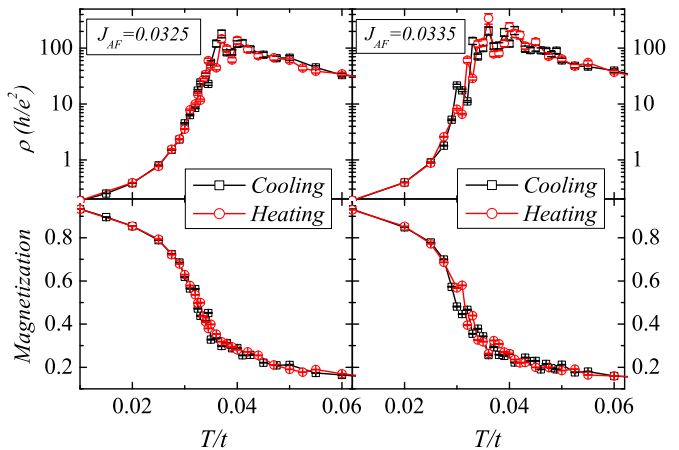}
\caption{(color online) Resistivity and uniform magnetization upon
cooling (black) and heating (red) for $J_{\rm AF}$=$0.0325$ (left
panel) and $J_{\rm AF}$=$0.0335$ (right panel). The result was
obtained on a $12\times12$ square lattice using $\lambda$=$1.2$,
density $n$=0.75, and in the clean limit ($\Delta$=$0$). No
hysteresis loop is observed in both quantities.} \label{F.PhT}
\end{center}
\end{figure}

Let us first revisit the thermal transition related to the CMR effect
that occurs in a narrow parameter regime
of the one-orbital model.\cite{Sen07} This transition is
simultaneously insulator to metal and PM to FM, upon cooling, similarly
as in experiments.
%
%
This MI transition and the associated magnetic transition, for a lattice
of size $12\times12$ and at electronic density $n$=$0.75$, are shown in
Fig.~\ref{F.PhT}. The critical temperature $T_{\rm C}$ of the magnetic
transition is approximately located at the resistivity peak, i.e.,
$T_{\rm C}\approx T_{\rm MI}$. At temperatures lower than
$T_{\rm MI}$ the resistivity falls down
rapidly, about two orders in magnitude within a narrow temperature
window $T/t \approx 0.01$. A similar
behavior in both the resistivity and the uniform magnetization is also
observed within a relatively narrow range of parameters: $\lambda\sim
1.1-1.2$ and $J_{\rm AF}\sim 0.02-0.034$.

In Fig.~\ref{F.PhT}, the resistivity and uniform magnetization phase
transitions were studied in the following way. For cooling, we
started the MC simulation at a high temperature $T\gg T_{\rm MI}$
using a random spin and phonon configuration. After MC converging at
the initial temperature, we started another simulation at a lower
temperature by using the final configuration from the previous
temperature as the new initial input. For heating, we proceeded in a
similar way, but the first simulation at the lowest temperature was
initialized with a perfect ferromagnetic spin configuration (note
that the CMR effect occurs very close to the FM-AF/CO transition but
slightly on the side of the FM phase, thus there is no need to start
any simulation with the AF/CO state). Both the heating and cooling
procedures were performed ``slowly'' in temperature. By this
procedure, we can detect possible different responses of the system
upon heating and cooling, which is usually associated with the
existence of first-order transitions. However, based on the results
shown in Fig.~\ref{F.PhT}, no hysteresis loop has been observed,
either in $\rho$ or in the magnetization. This suggests that both
the MI and magnetic transitions are of second-order in this
one-orbital model in the clean limit. Interestingly, our result is
consistent with a recent analytical work~\cite{salafranca} showing
that the transition is of second-order in the absence of quenched
disorder~\cite{SalafrancaBrey06}.



\subsection{Short-Range Spin and Charge Correlations \\
in the CMR Regime}
The CMR regime, where the resistivity peak develops during the MI
transition, is close to the bi-critical point separating the
FM and AF/CO states in the clean-limit one-orbital model phase diagram.~\cite{Sen07}
This indicates that at low temperatures, two competing ground states with different
long-range orders exist. Although by definition long-range order cannot survive
above a critical temperature, in this temperature regime
it is very common to find robust short-range correlations of the order
that will become stable at low temperatures. This certainly occurs in
the FM phase far from the region of competition with other states.
However, dramatically
different from this well-established concept, previous studies~\cite{Sen07}
observed that in the CMR regime, the short-range order that develops
above the Curie temperature is in the $charge$ sector. Namely, it is
{\it unrelated with the FM state}. Moreover, the observed
short-distance charge correlations are very similar to those of
the competing AF/CO state. Thus, the short-distance
physics above $T_{\rm C}$ near the bi-critical point is very different from the properties of
the  FM state, which is the actual ground state of the system.



The short-distance tendency toward a CO state above $T_{\rm C}$ and $T_{\rm MI}$
accounts for the
appearance of the resistivity peak at $T_{\rm MI}$. The system develops
CO charge correlation upon cooling and this proceeds together with an increase in
the resistance, since the AF/CO state is insulating.
In this section, we
confirm all these previous observations. More importantly, here
we will also consider the short-distance spin correlations above $T_{\rm C}$,
which were not analyzed before. We will arrive to the conclusion that
$both$ short-distance charge and spin correlations are important in
explaining the resistivity peak.

\begin{figure}[h]
\begin{center}
\includegraphics[
bbllx=10pt,bblly=10pt,bburx=590pt,bbury=860pt,%
     width=85mm,angle=0]{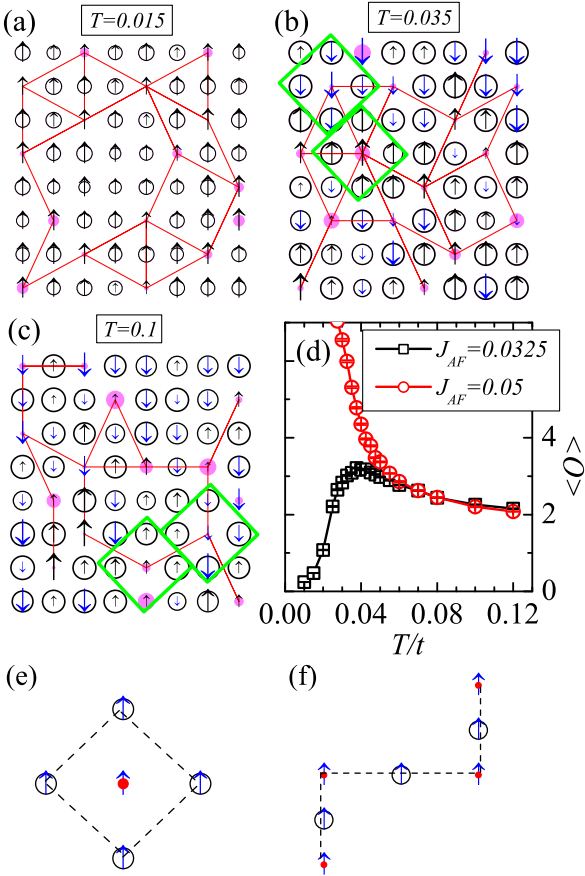}
\caption{(color online) (a)-(c) Typical MC snapshots of local charge
density and spin correlations on a $8\times8$ lattice with
$J_{\rm AF}=0.0325$, $n=0.75$, and $\lambda=1.2$ at three different
temperatures. In each snapshot, the circle radius is
proportional to the local charge density. The $16$ holes are represented
by filled circles. The length of the arrow at each site $i$ is
proportional to the spin-spin correlation $\langle \mathbf{S}_0\cdot
\mathbf{S}_i\rangle$, where $\mathbf{S}_0$ is a reference point
located on a hole site. Presenting this correlation, as opposed to directly showing the classical
spins, allows an easier visualization since now most spins are in the
plane of the figure.
Holes with $\sqrt{5}$ and $2$ correlations
are highlighted, as well as a couple of well-formed building blocks (see text) at the
two highest temperatures. (d) The temperature dependence of the proposed
operator $\hat{O}$ (Eq.~\ref{E.OrderParameter}). The
result of $\langle \hat{O}\rangle$ is averaged over $5\times10^4$ MC
steps. (e) The building block (see text) of the AF/CO state of the one-orbital
model at $n$=0.75. (f) Proposed building block of the AF/CO state in the
two-orbital model at $n$=0.5, for completeness.} \label{F.COSnapshots}
\end{center}
\end{figure}

An intuitive procedure to understand the short-distance
spin and charge correlations in the
one-orbital model is to analyze MC ``snapshots'', namely equilibrated configurations generated
by the MC  procedure. Examples are presented in Fig.~\ref{F.COSnapshots}.
Before analyzing them, and to guide the discussion, let us first consider what are the
expected ``building blocks'' in the charge-ordered state, and then we will investigate
if these building blocks are observed in the simulations.
Since it is anticipated that the short-distance behavior above $T_{\rm C}$
has strong similarities with the AF/CO competing state,
the expected building blocks, both for the one- and two-orbital models, are
shown in Figs.~\ref{F.COSnapshots} (e) and (f), respectively.
In the one-orbital model, the study in Ref.~\onlinecite{Sen07} has shown that
at $n$=0.75 the building block consists of a hole carrier surrounded by its
four neighboring sites, each carrying an electron. The five localized $t_{\rm 2g}$ spins involved
in this block
have the same orientation of those spins. This block has similarities with a FM polaron,
but quite contrary to a naive ``FM polaron gas'' scenario  the AF/CO
state is achieved by arranging these blocks $antiferromagnetically$ in a regular pattern
(see also Fig.~1(b)).
For the two-orbital model, shown here only for completeness, the most likely
building block
should be a small segment of the well-known CE-type FM zigzag chains (see Fig.~3(f)).

It is remarkable that the building blocks of the AF/CO state of the
one-orbital model can actually be observed in the MC snapshots shown in
Figs.~\ref{F.COSnapshots} (a) to (c), for a value of $J_{\rm AF}$
where the ground state is FM. We find that the short-range $2$ and $\sqrt{5}$ charge
correlations clearly exist at short distances, in both the high-temperature insulating PM and
low-temperature metallic FM phases. In the FM state, spins are
almost parallel to each other, and hole hopping is allowed. This
gives the FM state a metallic nature. Moreover, it will be shown below that
the charge is fairly mobile at low temperature, thus effectively the
building blocks loose their identity and they merge into a metallic FM state.
Different is the situation at larger temperatures.
When $T\gtrsim T_{\rm MI}$ (shown
as $T/t$=$0.035$ in Fig. \ref{F.COSnapshots}), the short-distance tendency to the competing
AF/CO state is significant. This tendency is
further confirmed by analyzing the short-range spin-spin correlations.
On one hand, the spins that are the nearest neighbors to a hole site are almost
parallel to the spin on that site, indicating short-range FM
correlations. On the other hand, spins localized on different hole sites
prefer to align antiferromagnetically at distances $\sqrt{5}$, as expected from the
discussion in Ref.~\onlinecite{Sen07}.
Then, the MC simulation clearly shows the presence of the building blocks described
in the previous paragraph (two of them are highlighted in Fig.~3(b)). As the temperature
cools down, these ``preformed'' building blocks rotate and orient themselves into
a FM pattern.
At higher temperature
$T/t$=$0.1$, both short-range spin and charge correlations are much
suppressed by thermal fluctuations, but the building blocks can still be
identified (see Fig.~3(c)).

By monitoring the above described MC snapshots, a simple
picture of the temperature
evolution of the system near the bi-critical point emerges: upon cooling from high temperature,
the building blocks develop, and first they start arranging in a pattern similar
to the AF/CO state. This causes the insulating properties above the Curie temperature.
Upon further cooling, the building blocks rotate
their relative orientation
from AF/CO to the FM state, rendering the state metallic.
To observe this effect more explicitly, we propose an operator $\hat{O}$
which characterizes the spin and charge correlations in the AF/CO
state, and study the temperature evolution of $\hat{O}$ in the
system. The operator is
\begin{equation}\label{E.OrderParameter}
\hat{O} = \sum_{\mathbf{i},\mathbf{j} = hole}
(-)^{\mathbf{i}-\mathbf{j}} \bar{\mathbf{S}}_\mathbf{i} \cdot
\bar{\mathbf{S}}_\mathbf{j} (1- n_\mathbf{i}) (1- n_\mathbf{j}),
\end{equation}
where $\bar{\mathbf{S}}_\mathbf{i} = \frac{1}{5} \left[
{\mathbf{S}}_\mathbf{i} + \sum_{\mathbf{\delta}}
{\mathbf{S}}_{\mathbf{i}+\mathbf{\delta}}\right]$ is the average of the
 classical spin
of the hole site~\cite{NoteHole} and its four neighbors
(namely, the ``spin block'' region enclosed by a diamond in
Fig.~\ref{F.GSsnap}(b) or~3(e)).
Note that $\langle \hat{O} \rangle$=$n^2_{\rm hole}$ in a
perfect AF/CO state with $n_{\rm hole}$ hole carriers. This same operator vanishes in
either a state with perfect FM order or in a PM state with uncorrelated
randomly distributed spins.

The temperature dependences of $\hat{O}$ for two values of $J_{\rm AF}$ are
shown in Fig.~\ref{F.COSnapshots}(d). At $J_{\rm AF}/t$=$0.05$, the
ground state does present long-range AF/CO order, and no resistivity peak was
observed. At this $J_{\rm AF}$ value, $\langle \hat{O}\rangle$ increases
monotonically upon cooling and saturates to a large finite value when $T\to
0$. The mean-value of this operator illustrates the transition
from PM to AF/CO states. Interestingly, in the CMR regime, such as for
$J_{\rm AF}/t$=$0.0325$ where the FM state is the ground state,
$\langle \hat{O}\rangle$ evolves fairly
differently. Upon cooling, it first slowly increases following very closely
the result at $J_{\rm AF}/t$=$0.05$, but then
develops a peak
at $T\approx T_{\rm MI}$, and subsequently drops fast to a nearly
vanishing value at the lowest
temperature investigated.
This non-monotonic temperature dependence clearly
reflects the enhanced tendency to the AF/CO state at $T\gtrsim
T_{\rm MI}$, intuitively observed in the MC snapshots. Remarkably, the
temperature dependence of $\langle \hat{O}\rangle$ and the
resistivity shows a qualitative similarity. It confirms that the
resistivity peak in the CMR regime is simply a reflection of the short-distance
tendency to the competing AF/CO state. We note that this idea has
been proposed in previous studies,\cite{Senetal06,Sen07} but only
the charge correlations were considered as relevant there. It is in this
work that  spin correlations are shown to be equally
important in understanding the CMR physics that appears in
the one-orbital model.
Charge and spin correlations develop together at short distances.

\begin{figure}[h]
\begin{center}
\includegraphics[
     width=90mm,angle=0]{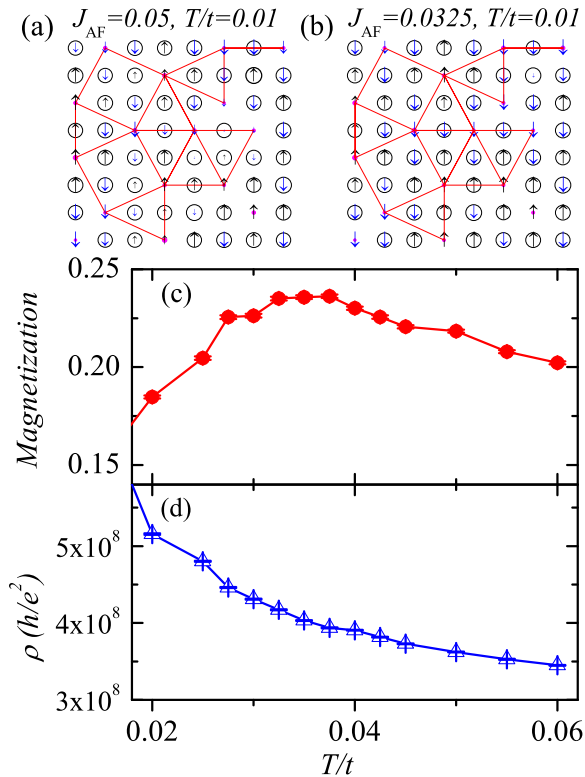}
\caption{(color online) MC results
 obtained by freezing the phononic degree-of-freedom
(see text for details). Snapshots of initial and final
configurations are shown in (a) and (b), respectively.
Also shown are the normalized magnetization (in
(c)) and resistivity (in (d)). With frozen charges, the DE mechanism cannot
produce the FM state at low temperatures.}
\label{F.freezephonon}
\end{center}
\end{figure}

To better understand the relation between charge and spin correlations, the
following MC simulation in the CMR regime, at $J_{\rm AF}$=$0.0325$ and
$\lambda$=$1.2$, was carried out.
The simulation was initialized with a configuration
where the system has charge correlations similar to the AF/CO state.
This was achieved by running
another simulation with $J_{\rm AF}$=$0.05$, $\lambda$=$1.2$, $T/t$=$0.01$,
and using  the final configuration as the initial configuration in the new run.
A snapshot of this initial
configuration is presented in Fig.~\ref{F.freezephonon}(a). During
the MC evolution, the oxygen lattice distortions degrees-of-freedom were kept $frozen$
to their initial
values, but the spins were allowed to change (note that we use one of the equilibrated
MC configurations as our ``frozen'' lattice and there the holes are not regularly distributed as in
a perfect CO state. However, the average over many MC configurations at the couplings used do give long-range order in the charge correlations).
By freezing the phonons,
the charge order is stabilized, and it provides a static
inhomogeneous background that is temperature independent.
If the spin and charge were
decoupled, when cooling the system in the CMR regime, spins can
still experience a PM-to-FM transition, and the system may be driven
to a metal due to the DE mechanism. However, as shown in
Figs.~\ref{F.freezephonon}(c) and (d), this is not what we obtained
from MC simulations. In fact, the system is never driven to a FM
state even at very low temperatures.
By observing the final configuration of the MC simulation at
$T/t$=$0.01$ in Fig.~\ref{F.freezephonon}(b) and comparing it with the
initial one, we find that spins are effectively also frozen
if the charges are frozen during the MC evolution. Thus,  the system is locked in the
AF/CO state. Interestingly, this picture holds even when $J_{\rm AF}$=$0$ (not shown).
Then, clearly spin flipping processes are greatly suppressed by freezing the
charge and lattice distortions. This suggests that: (i) the spin and charge degrees are
strongly coupled in the CMR regime; (ii) CMR cannot take place in a quenched
inhomogeneous charge background. The CMR is related to a more complex effect:
upon cooling across $T_{\rm C}$, the charge order partially melts allowing for
a metal to develop.


\subsection{Dynamical Properties of Charge and Spin Correlations in Monte Carlo Time}

In the one-orbital model without quenched disorder being analyzed here, the translation invariance symmetry
cannot be broken unless a state with long-range AF/CO
order is stabilized, and this occurs only at low temperatures and
sufficiently large $J_{\rm AF}$. Then, in the CMR regime where the resistivity peak appears,
the MC-time-averaged charge density
must  be the same at every site.
However, in the MC snapshots previously discussed, a short-distance charge localization resembling
the AF/CO state leads to a non-uniform distribution of charge $\langle n_i
\rangle$.
Then, as a state without long-range CO but containing
strong distortions resembling the AF/CO state, the charge localization
in the CMR regime can
only be understood in a dynamical context. In other words, even after the MC thermalization that
removes the dependence on the initial configuration, in the CMR regime the system is still inhomogeneous
at any instant during the simulation,
but it slowly becomes homogeneous when averaged over very long MC times.
It is, thus, interesting to investigate this dynamics. Note that, in principle, it is not obvious
that real-time and MC-time dynamics have any common  trends. However, both are based on local
events (local interactions in the real system to evolve in real time;
local updates in the MC evolution used here). Thus, the
expectation is that long-lived metastabilities and tendencies toward glassy behavior and a complex
landscape revealed via MC-time studies will have an analog in the real system.
To the extent that the analysis in MC time is
considered only qualitatively, our expectation is that crude
dynamical trends for the real system can be discussed based
on a local-update MC-time evolution.\cite{nonlocal}
With these caveats, here
several characteristic MC times, revealing dynamical correlations of the
spin and charge degrees-of-freedom in the system, will be numerically analyzed. It will be
argued that glassy properties emerge at the CMR regime, and we speculate (but do not prove rigorously)
that similar tendencies will occur in real manganites. A variety of experiments showing glassy
tendencies indicate that this assumption is realistic.\cite{glass}

\begin{figure}[h]
\begin{center}
\includegraphics[
bbllx=120pt,bblly=10pt,bburx=750pt,bbury=600pt,%
     width=70mm,angle=0]{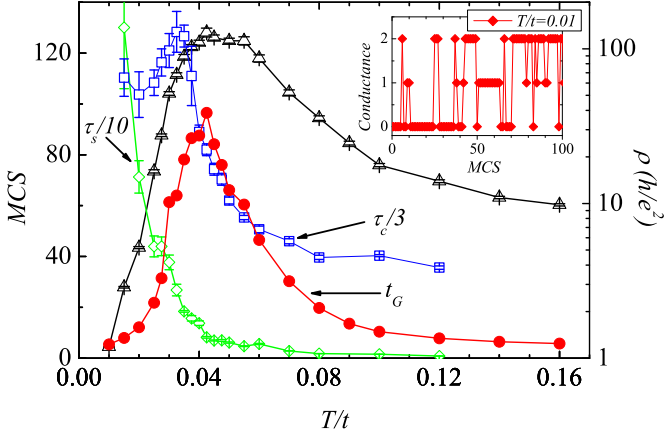}
\caption{(color online) Characteristic times that illustrate the spin and
charge MC dynamics in the one-orbital model are here shown as a function of temperature, together with the resistivity.
$t_{\rm G}$ is the average MC time that the conductance $G$ remains constant at one
of its quantized values. $\tau_{\rm c}$ is the autocorrelation time of the local charge
density. $\tau_{\rm s}$ is the autocorrelation time of the uniform spin structure factor.
The inset contains typical measurements of the conductance as a function of MC
time, illustrating the discrete values it takes.} \label{F.COdynamics}
\end{center}
\end{figure}

Consider first the
MC-time evolution of the conductance (inset of
Fig.~\ref{F.COdynamics}). In the present type of studies based on calculating
the quantum-mechanical transmission to obtain conductances, it is well-known that
the conductance is
quantized, namely it varies among discrete values due to the presence
of an integer number of channels of transmission, from one side to the other of the
cluster investigated. The observed ``jumps'' between two
quantized values reflects on changes that occurred in the charge and spin correlations. Thus,
the average MC time $t_{\rm G}$ that the conductance remains at a certain
quantized value provides information about the system MC dynamics. From
Fig.~\ref{F.COdynamics}, it is interesting to observe that $t_{\rm G}$
has a very clear peak precisely at the metal-insulator transition temperature. In other
words, as the system is cooled down, concomitant with the development of short-range
charge and spin correlations, a variety of metastable states develop, as in a glass. The system tends to
be trapped in those states and evolving to other states is difficult. Eventually
at very low temperatures, deep into the FM state, the metastabilities vanish and the
system rapidly evolves in MC time. It is remarkable that $t_{\rm G}$
qualitatively follows the shape of the resistivity $\rho$. This suggests that the charge localization close
to $T_{\rm MI}$ has glassy characteristics,\cite{glass} and if the free energy were available in the CMR
regime, it would show a complex landscape of peaks and valleys.
The long lifetimes near the CMR resistivity peak show the practical difficulty in achieving true translation invariance:
for this to occur, the system must visit dynamically all the many competing
arrangements of charge and spin to render the average uniform.

We can also study an autocorrelation time $\tau_{\rm c}$ related with the
charge. For this purpose, let us define the autocorrelation function $A(t)$ of
the local charge density as
\begin{equation}\label{E.ChargeAutoCorr}
A(t) = \frac{\frac{1}{NT_0}\sum_{i=1}^N\sum_{t'=1}^{T_0} [ n_i^{t+t'}-
\bar{n}_i] [ n_i^{t'}- \bar{n}_i]}
{\frac{1}{NT_0}\sum_{i=1}^N\sum_{t'=1}^{T_0} [ n_i^{t'}- \bar{n}_i] [
n_i^{t'}- \bar{n}_i]},
\end{equation}
where $\bar{n}_i = \frac{1}{T_0}\sum_{t=1}^{T_0} n_i^t$, and $T_0$ is the total number of MC steps used
in the measurements (50,000 in this case). It was observed that $A(t)$ decays
exponentially as $A(t)\sim e^{-t/\tau_{\rm c}}$, where $\tau_{\rm c}$ is the
autocorrelation time of the local charge density. The temperature
dependence of $\tau_{\rm c}$ is also shown in Fig.~\ref{F.COdynamics}. It has a
similar shape as $t_{\rm G}$, indicating that charge has a tendency toward localization
in long-lived states near the CMR peak.
However, the peak in $\tau_{\rm c}$ is
shifted to temperatures lower than the metal-insulator transition temperature.
Thus, charge localization is
still quite stable even below $T_{\rm MI}$.
This picture changes at very
low temperature, where $\tau_{\rm c}$ increases again, and the system
stabilizes a metallic state with uniform charge distribution. This increase of $\tau_{\rm c}$
at low temperature is unrelated to charge localization, but it reflects on the stable uniformity
of the FM state charge distribution.

As discussed above, spin dynamics is also important in
understanding the resistivity peak in the CMR regime. Similar to
charges, an autocorrelation time can be defined for the spins. This
autocorrelation function reads
\begin{equation}
A(t) = \frac{\sum_{t'=1}^{T_0} [ S^{t+t'}- \bar{S}] [ S^{t'}-
\bar{S}]} {\sum_{t'=1}^{T_0} [ S^{t'}- \bar{S}] [ S^{t'}- \bar{S}]},
\end{equation}
where $S = \frac{1}{N} \sum_{i,j} \mathbf{S}_i \cdot \mathbf{S}_j$
is the uniform spin structure factor, and $\bar{S} = \frac{1}{T_0}\sum^{T_0}_{t=1}
S$. The autocorrelation time $\tau_{\rm s}$ is then defined through
$A(t)\sim e^{-t/\tau_{\rm s}}$. As shown in Fig.~\ref{F.COdynamics}, it
increases quickly at $T<T_{\rm MI}$ indicating the stabilization of a FM
state.

\begin{figure}[h]
\begin{center}
\includegraphics[
     width=80mm,angle=0]{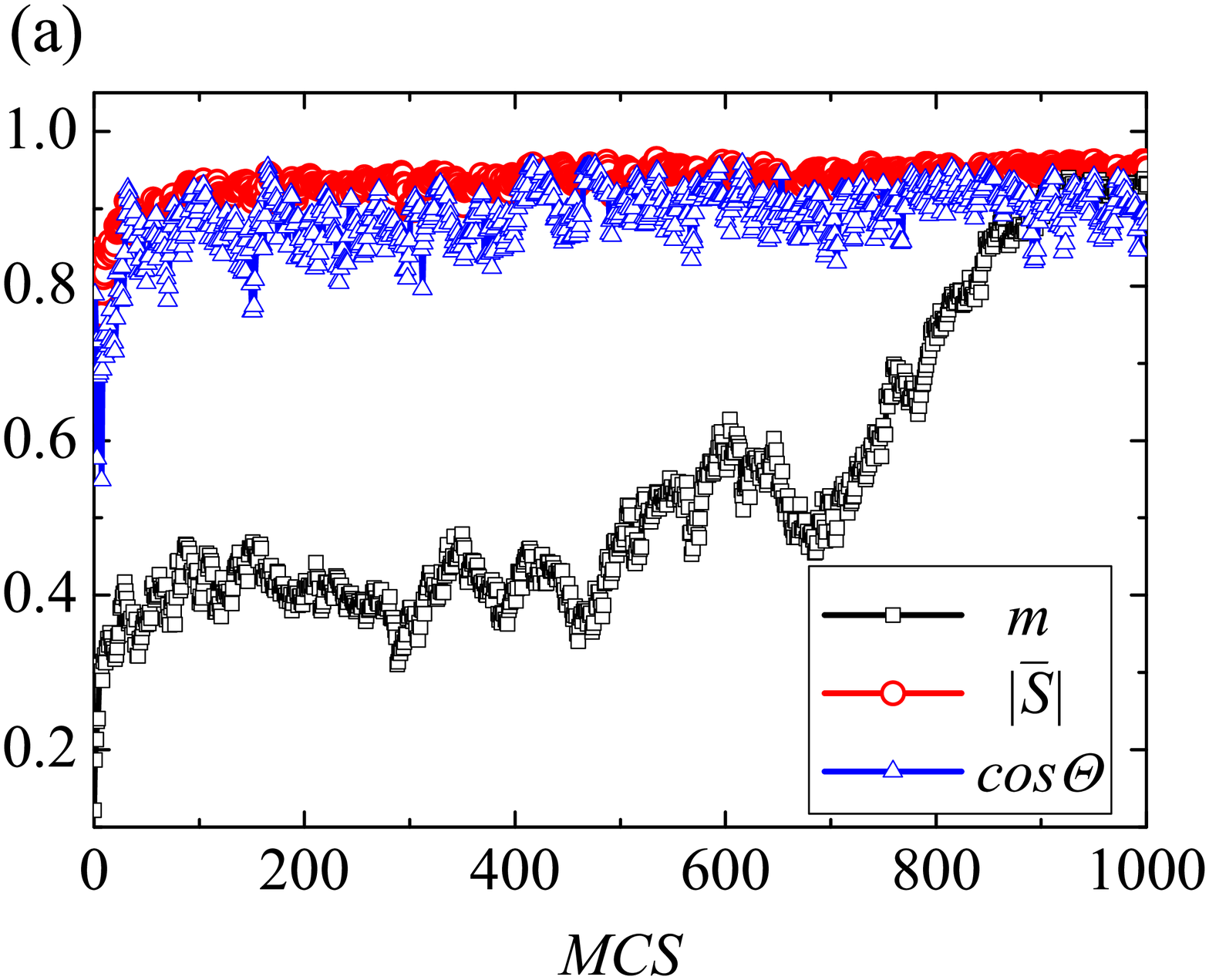}
\vskip -0.6cm
\includegraphics[
bbllx=40pt,bblly=10pt,bburx=587pt,bbury=560pt,%
     width=80mm,angle=0]{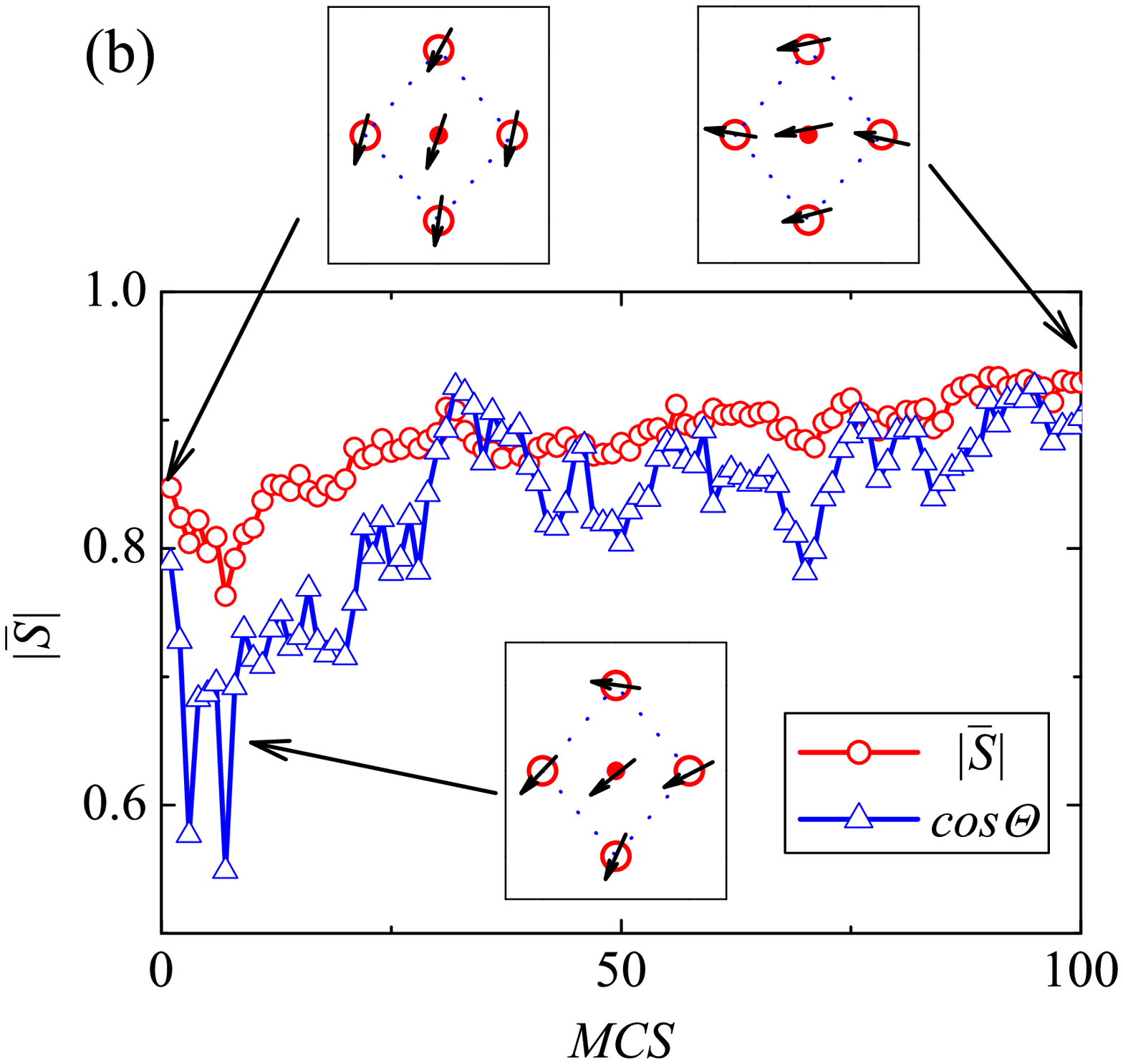}
\caption{(color online) (a) MC time evolution of $|\bar{S}|$, $\cos
\Theta$, and the normalized magnetization $m$. The MC simulation is
initialized by a configuration with short-distance AF/CO order
(similar to that shown in Fig.~\ref{F.GSsnap}(b)), but the simulation occurs
at very low temperature, where a FM ground state is stable (similar to that
shown in Fig.~\ref{F.GSsnap}(a)). (b) MC-time evolution of $|\bar{S}|$
and $\cos\Theta$, at shorter MC-time scales than in (a).
Three snapshots of a typical
block of spins in the early stages of MC evolution are also shown.} \label{F.SpinBlock}
\end{center}
\end{figure}

We found that very short-range FM correlations exist even in the AF/CO
state. This is clear by observing a typical MC snapshot of spin
correlations. In a perfect AF/CO state, each hole is surrounded by
four neighbors that have their classical spins parallel to the classical spin on that hole
site. As previously discussed, such spin blocks are then antiferromagnetically arranged in
the AF/CO state (see Fig.~\ref{F.GSsnap}(b)). However, if all these
block spins align, a FM state is reached. It is then
interesting to ask how the spins rotate when experiencing a transition
from the short-distance AF/CO state in the CMR regime to the FM state stable at lower
temperatures. Do they rotate coherently, behaving
as blocks of spins, or the spins rotate more individually one at a time?
To clarify this matter, let us study the MC evolution of the average block
spin $|\bar{\mathbf{S}}| = \sqrt{\frac{1}{N}\sum_i
\bar{\mathbf{S}}_i^2}$. Let us also study the average angle between the spin at a hole
site and its four nearest-neighboring spins, defined as  $\cos \Theta =
\frac{1}{(1-n)L^2} \sum_{i=holes} \mathbf{S}_i \cdot
\mathbf{S}_{i+\hat{x}}$.
The results are shown in Fig.~\ref{F.SpinBlock}. The MC simulation was
performed at low temperature where a FM ground state is finally
reached. However, it was started with an initial configuration
obtained in an almost perfect AF/CO state.\cite{NoteConfig} The
initial and final configurations have very similar values for
$|\bar{\mathbf{S}}|$. If the spins rotate incoherently,
$|\bar{\mathbf{S}}|$ should be first reduced substantially and then grow back when a large portion
of spins are parallel (to establish the long-range FM order), and
$\cos \Theta$ should be randomly centered around $0$. If, on the other
hand, spins rotate coherently, both $|\bar{\mathbf{S}}|$ and $\cos
\Theta$ should remain nearly constant and close to $1$ over the MC time evolution. As seen
from Fig.~\ref{F.SpinBlock}(a), both quantities converge to their
equilibrium values very fast, even before the thermal equilibrium of
the FM state is established. It is interesting that both
quantities do not change too much during the MC time evolution, which indicates that spins rotate
coherently in blocks, at least as a first approximation.

These results would suggest the following picture: although not perfectly,
the spins in a block mainly rotate coherently, such that they keep parallel
to one another during the MC evolution. This picture is consistent with the
charge localization scenario for the AF/CO state, and the MI
transition in the CMR regime can be understood in the following way.
At $T>T_{\rm MI}$, holes are localized in these local FM spin blocks,
and different spin orientations between adjacent blocks suppress
inter-block charge transfer, giving the PM phase an insulating
nature. However, holes are not frozen at their initial sites. As
they travel from site to site, they coherently rotate spins on their
neighboring sites in the block. At $T$$<$$T_{\rm c}\thickapprox T_{\rm MI}$,
long-range spin correlations establish.  Blocks are aligned in
parallel, charges are no longer localized in blocks but can freely
move all over the lattice, developing metallic properties.
Note the consistency between the above described picture and
the ``correlated polaron" picture often mentioned in the
experimental literature\cite{correlated-polarons} by referring to the local FM blocks
around holes as ``small polarons''. These polaronic entities are $not$ independent
but they have tendencies to very particular short-distance arrangements
in the CMR regime. This agreement between
theory and experiment also supports the view
that at least qualitatively the study of
MC dynamics provides useful information regarding
the actual real time dynamics of manganite compounds.


\section{The local density-of-states
in the one-orbital model}
\begin{figure}[h]
\begin{center}
\includegraphics[
bbllx=160pt,bblly=10pt,bburx=750pt,bbury=600pt,%
     width=65mm,angle=0]{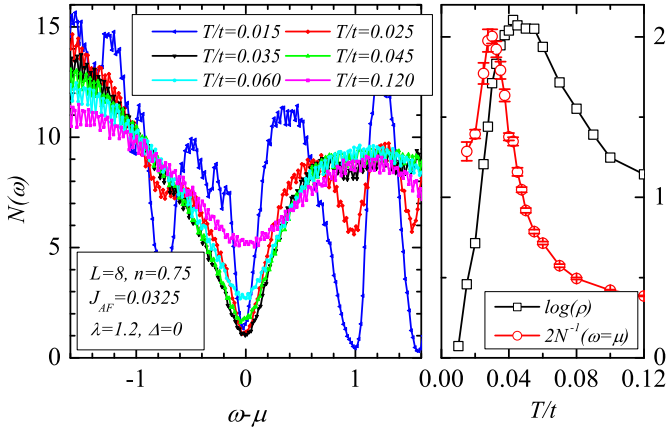}
\caption{(color online) {\it Left:} Density of states $N(\omega)$ at
various temperatures, in the CMR regime. A pseudogap develops in both the PM insulating
and FM metallic states. {\it Right:} temperature evolution of
$N^{-1}(\omega=\mu)$ and the resistivity $\rho$.} \label{F.Now}
\end{center}
\end{figure}

Motivated by recent STS experiments,\cite{Seiroetal07,Singhetal07}
we have systematically studied the behavior of the LDOS in the one-orbital model
for manganites, with focus on the CMR regime. For this analysis, we have worked
on an $8\times 8$ lattice with model
parameters $J_{\rm AF}$=$0.0325$ and $\lambda$=$1.2$. Before proceeding with
the local DOS analysis, note
that the site averaged LDOS is just the DOS $N(\omega)$ of the
system. Unlike what the experimental results have suggested recently,\cite{Seiroetal07} we do not
observe any $hard$ gapped feature in $N(\omega)$, at any
temperature. However, as shown in Fig.~\ref{F.Now}, in the CMR
regime of the one-orbital model, our results for the DOS present a clear and fairly deep PG
centered at the chemical
potential, both below and above the critical temperatures. Starting
from the high-temperature insulating phase, the DOS at the chemical
potential first decreases upon cooling the system, it reaches a minimum
at approximately $T_{\rm MI}$, and then grows up again. The temperature dependence of
the inverse of the DOS, $N^{-1}(\omega=\mu)$, qualitatively follows the shape of the
resistivity. An enhancement in $N^{-1}(\omega=\mu)$ reveals a clear
tendency to an insulating state, result compatible with all the previous
analysis of transport and spin/charge short-distance correlations.
Once spatial
fluctuations are included, most features observed in the DOS apply to LDOS as well.

Although a hard gap in the LDOS was not observed in our model even
in the insulating phase, we found several similarities between
properties of the PG regime and those of a normalized conductance curve
reported in experiments.\cite{Seiroetal07} The obvious one is that the PG exists in
the full CMR temperature range we studied, both in the
FM metallic state and in the PM insulating phase, while in experiments a hard gap is
found in both regimes as well.
Moreover, in the experiment,
the width of the LDOS gap near the chemical potential $\mu$ was determined. Its temperature
dependence was associated with the development of a polaron binding
energy. For further comparison with experimental results, we have studied
the width of the PG in our LDOS. The local width $\Delta E_i$ is determined by
taking the half-distance between the two peaks that are the closest to the chemical
potential, at
each lattice site $i$. The site-averaged width $\Delta E$, its
variance $\sigma_{\Delta E}$, and the distribution of $\Delta E_i$,
denoted by $P(\Delta E_i)$, are then calculated.

\begin{figure}[h]
\begin{center}
\includegraphics[
bbllx=100pt,bblly=10pt,bburx=750pt,bbury=600pt,%
     width=75mm,angle=0]{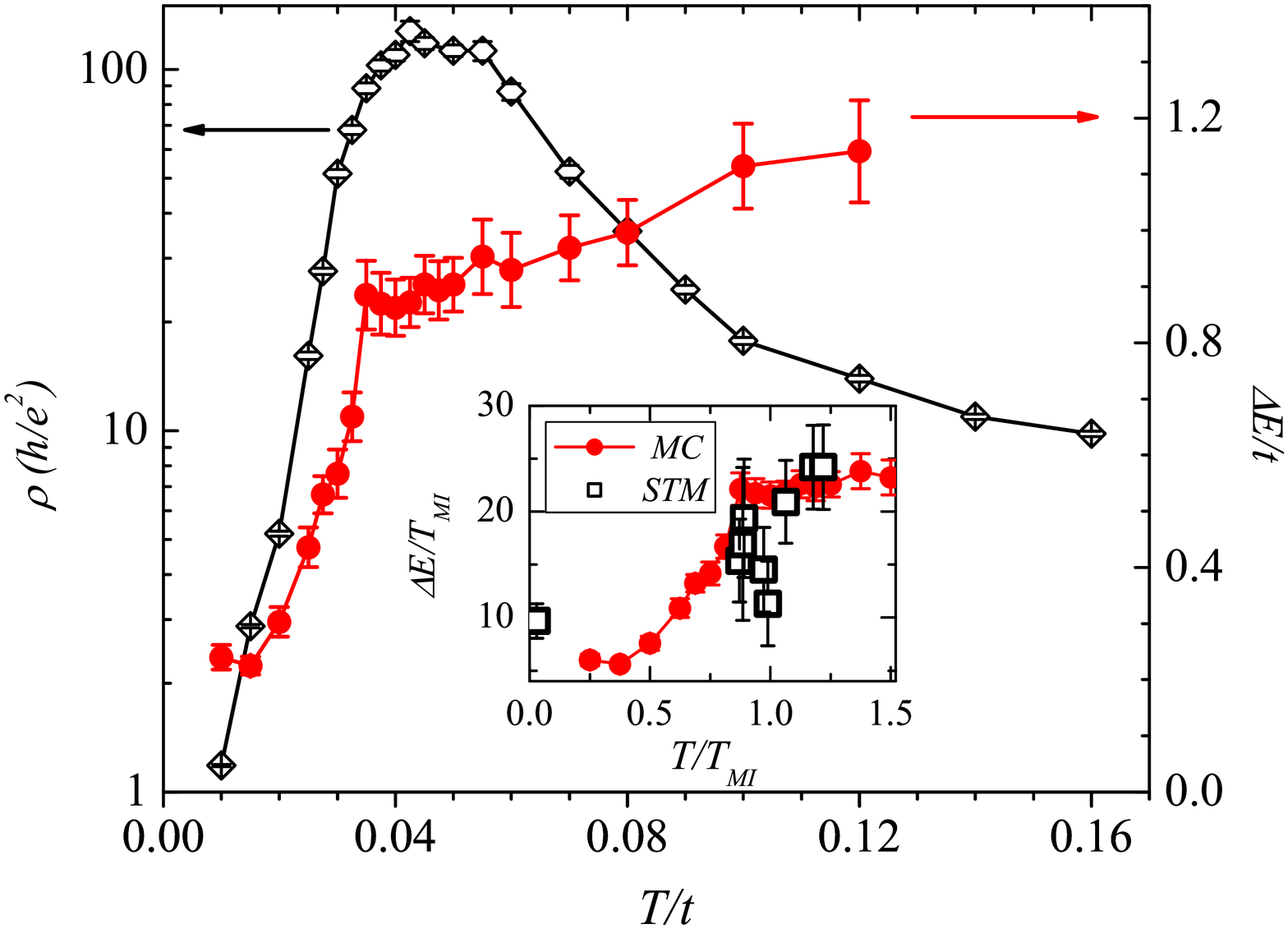}
\caption{(color online) Temperature dependence of the site-averaged
width $\Delta E$ of the pseudogap in the LDOS. Also shown is the
resistivity (black diamonds) at the same model parameters. {\it Inset:}
$\Delta E$ rescaled by the critical temperature of the MI transition
$T_{MI}$. Also shown as black squares are the experimentally determined ``polaronic gaps''
(extracted from Ref.~\onlinecite{Seiroetal07}) rescaled by $T_{\rm MI}$.
} \label{F.PseudoGapWidth}
\end{center}
\end{figure}

In Fig.~\ref{F.PseudoGapWidth}, the site-averaged width $\Delta E$ is
shown together with the resistivity. $\Delta E$ has a small
value at low temperatures in the FM metallic phase, as expected. However, this
quantity increases
upon raising the temperature. Close to $T\sim T_{\rm MI}$, $\Delta E$
develops a cusp and it flattens into
a plateau at about where the resistivity shows a peak.
Actually, at exactly $T\approx T_{\rm MI}$ it reaches a local minimum, namely a
little dip on the plateau, but it is difficult to judge if such a small
effect can survive the bulk limit. Upon further heating, $\Delta E$ slowly
increases again. To directly compare
our results with experimental data, we rescaled $\Delta E$ and
the temperature $T$ with respect to $T_{\rm MI}$. The rescaled dimensionless
$\Delta E$ together with
experimental data (rescaled as well) are shown in the inset of
Fig.~\ref{F.PseudoGapWidth}. The quantitative
agreement is very good in almost the
full temperature range.
Nevertheless, there is a difference between experiments and our
calculation close to $T\approx T_{\rm MI}$. Experimentally
a notorious dip was observed, but this feature appears to be very weak in our simulations,
as already discussed. However, the overall behavior is at least qualitatively the same
both in theory and experiments.
\begin{figure}[h]
\begin{center}
\includegraphics[
bbllx=50pt,bblly=10pt,bburx=614pt,bbury=882pt,%
     width=75mm,angle=0]{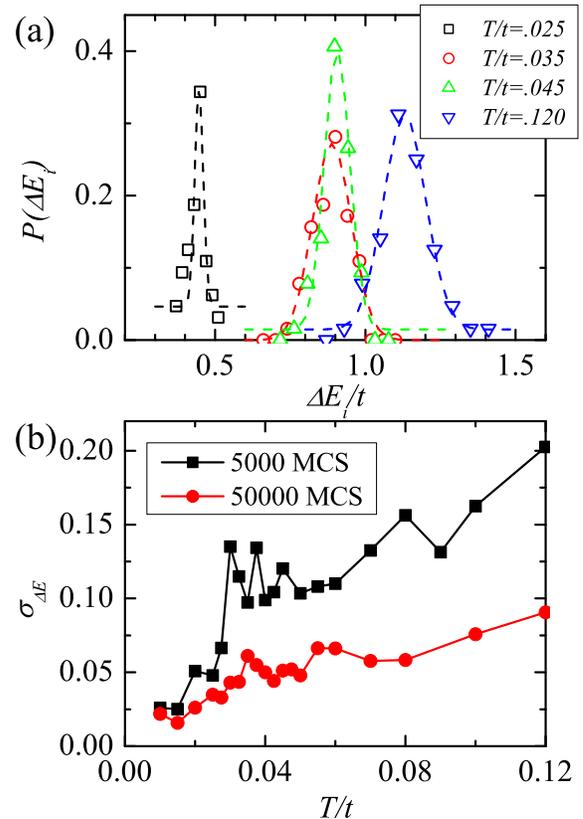}
\caption{(color online) Statistics of the width of the LDOS pseudogap
$\Delta E_i$. (a) A general Gaussian probability distribution of
$\Delta E_i$ at various temperatures in both the FM metallic and PM
insulating phases is observed, when MC-averaged over long times.
(b) The standard deviations of $\Delta E_i$. The
system behaves more inhomogeneously in shorter MC-time scales.}
\label{F.GapWidthStatistics}
\end{center}
\end{figure}

As a further comparison, we present the distribution of LDOS widths,
$P(\Delta E)$, in the PM insulating phase, as well as in the FM
metallic phase. As shown in Fig.~\ref{F.GapWidthStatistics}(a),
we have observed that in both phases the distribution can be fitted by a
Gaussian, in full agreement with the experimental findings.\cite{Seiroetal07} At first
glance, a single-peak Gaussian distribution implies that the system is
homogeneous, with no competing metallic and insulating domains.
However, we have to emphasize that the distribution in our model is
taken from very long MC simulations, similarly as
the STS experiment which is
``slow'' in typical time scales of relevance for electronic systems.
Thus, the Gaussian distribution of widths indicates that the
system is statistically homogeneous when time averaged. In other words, the
site-averaged LDOS PG width, $\Delta E$, equals to the PG width of  the
DOS in a very long MC run. But it does not exclude the possibility
of having a dynamical inhomogeneity, such as the charge localization and
formation of local FM spin blocks discussed in the previous sections. Since
the distribution of the PG width is Gaussian, its variance
$\sigma_{\Delta E}$ is then a measure of the inhomogeneity in the
system. This quantity is illustrated in Fig.~\ref{F.GapWidthStatistics}(b):
$\sigma_{\Delta E}$ strongly depends on the number of MC steps in
the simulation, especially close to $T_{\rm MI}$ in the insulating
phase, where the dynamical charge localization and correlations are
enhanced. If the number of MC steps for the MC measurements is
comparable to the characteristic time of the system dynamics, such
as the autocorrelation times $\tau_{\rm c}$ and $\tau_{\rm s}$, a different
distribution of LDOS PG widths may appear.

\begin{figure}[h]
\begin{center}
\includegraphics[
bbllx=0pt,bblly=0pt,bburx=612pt,bbury=794pt,%
     width=85mm,angle=0]{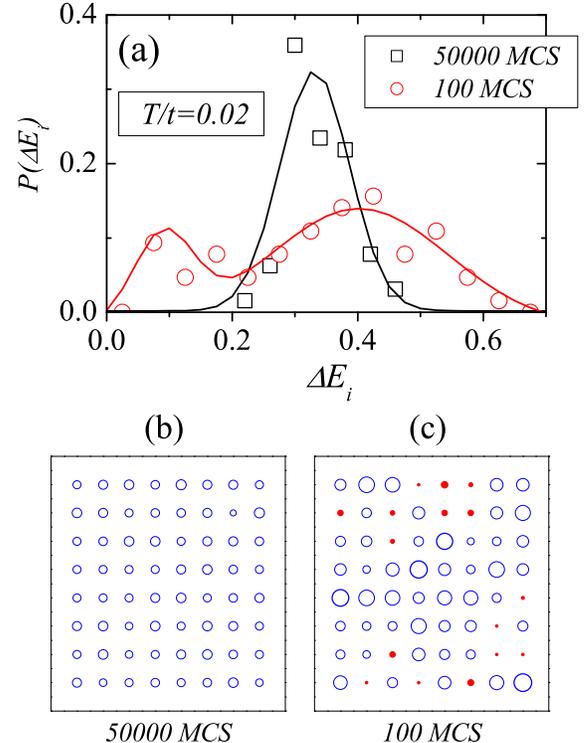}
\caption{(color online) (a) Distribution of the width of the LDOS
pseudogap $\Delta E_i$ at different MC time scales. Real space
images of $\Delta E_i$ measured under $50,000$ and $100$ MC steps at
$T/t=0.02$ are also provided in (b) and (c), respectively. The
radius of each circle is proportional to the local $\Delta E_i$
value at that site. Open circles corresponds to $\Delta E_i/t
\geqslant 0.2$, whereas solid (red) ones corresponds to $\Delta E_i/t <
0.2$.} \label{F.GapWidthIns}
\end{center}
\end{figure}

Following this idea, we
compare the distribution of $\Delta E_i$ between two MC simulations.
These simulations are performed for the same couplings and other parameters,
but one is done with
$5\times 10^4$ MC steps, while the other one has only $100$ MC steps
(both after $10^5$ MC steps for thermalization). As
shown in Fig.~\ref{F.GapWidthIns}(a), the distribution is now quite
different: a Gaussian is observed for the long run, but a
double-peak distribution, which can be fitted by a superposition of
two Gaussian's, applies to the short run. The real space images in
Fig.~\ref{F.GapWidthIns}(b) and (c) further confirm that the system
looks more inhomogeneous at very short time scales. For the same
reason, if one were able to perform the STS experiment in a much
shorter time scale than currently possible, namely with times
comparable to characteristic relaxation times in the system, then a
different gap distribution revealing the intrinsic dynamical
inhomogeneity would be observed in the experiment.

It is important to further discuss this surprising agreement between our
calculation and the STS experiments. While it is easy to imagine a gapped
spectrum in an insulator, finding a gap in the metallic phase seems very
unusual. The explanation proposed in Ref.~\onlinecite{Singhetal07} is clearly
a possibility, namely the surface behaves differently than the bulk.
However, if the concept of a hard gap is replaced by the
pseudogap, then another possible scenario appears.
A PG in the metal renders the system a ``bad metal'', as observed experimentally since
the values of the resistivity are high in the low temperature phase.
Finding charge approximately
localized and correlated in the FM metallic state near the MI transition is not
too surprising, according to the
physical picture and results we have discussed in previous sections.
It is then natural to explain the STS gapped feature of the  conductance
spectrum observed experimentally as a result of a fairly deep PG, which
cannot be distinguished experimentally from a hard gap. In retrospect this
is fairly obvious: a hard gap cannot exist in a metal.
But for a PG coexisting with bad metallic properties, there is no contradiction.


\section{Conclusion}

In conclusion, we have investigated the MI transition in the CMR regime,
and the associated PM to FM transition, using the one-orbital model with
cooperative oxygen lattice distortions and superexchange coupling $J_{\rm AF}$, at
electronic density $n=0.75$. Both the resistivity and magnetization
are continuous as a function of temperature, indicating a second order phase transition.
However, strong tendencies to develop spin and charge short-range order were observed. It
has been shown
that the tendency toward the
AF/CO state at short-distances accounts for the appearance of the sharp resistivity
peak at the critical Curie temperature. Robust charge
localization and correlations exist in both the insulating
and metallic phases, and are enhanced in the temperature range where
a resistivity peak exists. Charge localization together with
correlations in the spin and charge degrees-of-freedom are shown to be dynamical, at
least with respect to the MC time in the simulations,
which can be characterized by autocorrelation times $\tau_{\rm c}$ and
$\tau_{\rm s}$. The model studied here presents properties in the CMR regime
that resemble a glassy system.

We have also studied the LDOS of the one-orbital model in the CMR regime. A
PG is observed at all temperatures. The width of this PG, together
with the variance and distribution of the width, are calculated and
compared with results from a recent STS experiment. To our
surprise, given the crudeness of the calculations, theory and experiments
agree not only qualitatively, but in some cases even quantitatively. This study
provides an explanation to the puzzling STS experimental results, suggesting
that the gapped features in the normalized conductance spectrum
observed in the STS experiment actually originates from a pseudogap. We also point
out that the observed homogeneity in STS investigations
may reflect on the time-averaged properties of
the system. The system is actually dynamically inhomogeneous, at least
in our Monte Carlo time evolution. Our prediction is that
experimental techniques that are fast, namely with time scales comparable
to the system's intrinsic dynamics in the CMR regime,  should see indications of
the widely expected nanoscale phase separation. However, slow techniques, such as STS,
should observe an homogeneous state in the same regime.

\section{Acknowledgement}

We thank S. Seiro and S. Yunoki for useful discussions. Work
supported by the NSF grant DMR-0706020 and the Division of Materials Science and
Engineering, U.S. Department of Energy, under contract with UT-Battelle, LLC. A portion of this research at Oak Ridge
National Laboratory's Center for Nanophase Materials Sciences was sponsored by the Scientific User Facilities
Division, Office of Basic Energy Sciences, U.S. Department of Energy.
S.~D. is also supported by the China Scholarship Council (2007U03040).


\begin{thebibliography}{99}
\bibitem{SalafrancaBrey06} 
According to Ref.~\onlinecite{salafranca}, the transition should
turn into first order when the strength of quenched disorder exceeds
a certain critical value. In future work, it will be interesting to
study the MI transition in the model with quenched disorder, and
numerically check the validity of this analytical claim.
\bibitem{NoteHole} In practice, hole sites are defined as sites with
the lowest $(1-n)L^2$ local densities of states at each MC step,
where $n$ is the number of electrons in a $L\times L$ system.
\bibitem{nonlocal} If the MC evolution were carried out using non-local updates (using, e.g.,
cluster updates), then the analogy between the real system and the MC evolution is lost.
\bibitem{NoteConfig} The simulation is performed at $T/t=0.01$ with
$J_{\rm AF}=0.0325$ and $\lambda=1.2$, where a metallic FM state is
finally stabilized. However, in this simulation the initial
configuration is prepared by taking the final configuration of
another long MC simulation in the AF/CO state (with $J_{\rm
AF}=0.05$, $\lambda=1.2$, and $T/t=0.01$).
\end{thebibliography}
\end{document}